\documentclass[%
reprint,
superscriptaddress,
showpacs,preprintnumbers,
aps,
pra,
]{revtex4-1}

\usepackage[dvipdfmx]{graphicx, color}
\usepackage{mediabb}
\usepackage{amsmath}
\usepackage{color}
\usepackage{bm}
\usepackage{mathrsfs}
\usepackage[hyperfootnotes=false]{hyperref} 
\usepackage{mathtools} 
\usepackage{braket}
\usepackage{amssymb}
\usepackage[normalem]{ulem} 
\usepackage{xcolor,cancel} 
\usepackage{dcolumn}
\usepackage[multiple]{footmisc} 

\newcommand{\Apeak}{A_{\mathrm{peak}}}

\usepackage{amsmath}	
\begin{document}

\title{Trajectory analysis of high-harmonic generation from periodic crystals}


\author{Takuya Ikemachi}
  \email[]{ikemachi@gono.phys.s.u-tokyo.ac.jp}
  \affiliation{Department of Physics, Graduate School of Science, The University of Tokyo, 7-3-1 Hongo, Bunkyo-ku, Tokyo 113-0033, Japan}
\author{Yasushi Shinohara}
  \affiliation{Photon Science Center, Graduate School of Engineering, The University of Tokyo, 7-3-1 Hongo, Bunkyo-ku, Tokyo 113-8656, Japan}
\author{Takeshi Sato}
  \affiliation{Photon Science Center, Graduate School of Engineering, The University of Tokyo, 7-3-1 Hongo, Bunkyo-ku, Tokyo 113-8656, Japan}
  \affiliation{Department of Nuclear Engineering and Management, Graduate School of Engineering, The University of Tokyo, 7-3-1 Hongo, Bunkyo-ku, Tokyo 113-8656, Japan}
\author{\\Junji Yumoto}
  \affiliation{Department of Physics, Graduate School of Science, The University of Tokyo, 7-3-1 Hongo, Bunkyo-ku, Tokyo 113-0033, Japan}
  \affiliation{Institute for Photon Science and Technology, Graduate School of Science, The University of Tokyo, Tokyo, 113-0033 Japan}
\author{Makoto Kuwata-Gonokami}
  \affiliation{Department of Physics, Graduate School of Science, The University of Tokyo, 7-3-1 Hongo, Bunkyo-ku, Tokyo 113-0033, Japan}
\author{Kenichi L. Ishikawa}
  \affiliation{Photon Science Center, Graduate School of Engineering, The University of Tokyo, 7-3-1 Hongo, Bunkyo-ku, Tokyo 113-8656, Japan}
  \affiliation{Department of Nuclear Engineering and Management, Graduate School of Engineering, The University of Tokyo, 7-3-1 Hongo, Bunkyo-ku, Tokyo 113-8656, Japan}


\date{\today}

\begin{abstract}

We theoretically study high-harmonic generation (HHG) from solids driven by intense laser pulses using a one-dimensional model periodic crystal.
By numerically solving the time-dependent Schr\"{o}dinger equation directly on a real-space grid, we successfully reproduce experimentally observed unique features of solid-state HHG such as the linear cutoff-energy scaling and the sudden transition from a single- to multiple-plateau structure.
Based on the simulation results, we propose a simple model that incorporates vector-potential-induced intraband displacement, interband tunneling, and recombination with the valence-band hole.
One key parameter is the valley-to-peak amplitude of the pulse vector potential, which determines the crystal momentum displacement during the half cycle.
When the maximum peak-to-valley amplitude $\Apeak$ reaches the half width $\frac{\pi}{a}$ of the Brillouin zone with $a$ being the lattice constant, the HHG spectrum exhibits a transition from a single- to multiple-plateau structure, and even further plateaus appear at $\Apeak = \frac{2\pi}{a}, \frac{3\pi}{a}, \cdots$.
The multiple cutoff positions are given as functions of $\Apeak$ and the second maximum $\Apeak^{\prime}$, in terms of the energy difference between different bands.
Using our recipe, one can draw electron trajectories in the momentum space, from which one can deduce, for example, the time-frequency structure of HHG without elaborate quantum-mechanical calculations.
Finally, we reveal that the cutoff positions depend on not only the intensity and wavelength of the pulse, but also its duration, in marked contrast to the gas-phase case.
Our model can be viewed as a solid-state and momentum-space counterpart of the familiar three-step model, highly successful for gas-phase HHG, and provide a unified basis to understand HHG from solid-state materials and gaseous media.

\end{abstract}


\maketitle

\section{Introduction}
Advances in ultrashort  intense laser techniques have paved the way to investigate strong-field and attosecond physics.
In particular, high-harmonic generation (HHG) from gas-phase atoms and molecules has been one of the main targets of research for three decades, which has led to successful applications such as attosecond pulse generation \cite{Goulielmakis2008,Zhao_2012} and coherent keV x-ray sources \cite{Popmintchev2012,Kfir_2014}
as well as powerful means to observe and manipulate ultrafast electron dynamics \cite{Itatani_2004,Kling2006,Uiberacker2007,Cavalieri2007,Smirnova_2009,W_rner_2010,Schultze2013,Schultze2014}.

Solid-state materials have recently emerged as a new stage of strong-field and attosecond physics.
Stimulated by the discovery by Ghimire {\it et al.} \cite{Ghimire2011} and subsequent successful observations \cite{Schubert_2014,Luu2015a,Vampa2015,Hohenleutner2015a,Ndabashimiye_2016}, the mechanism of HHG from solids (we focus on crystalline dielectrics and semiconductors) are under intensive discussion \cite{Golde2008,Muecke2011,Ghimire2012,Otobe2012,Korbman2013,Hawkins2013,Higuchi2014a,Vampa2014,Schubert_2014,Vampa2015,Luu2015a,Hohenleutner2015a,Vampa2015a,Vampa2015b,McDonald2015,Hawkins2015a,Wu2015,Tamaya2016,Guan2016,Ndabashimiye_2016}.
Intense laser fields are generally considered to induce both interband and intraband electron dynamics in the momentum space in solids;
the former refers to (usually vertical, tunneling) transitions between different bands, and the latter to displacements in the $k$ space within one band.
Early work focused on the intraband dynamics \cite{Ghimire2011,Schubert_2014,Luu2015a,Hawkins2013,Hawkins2015a}.
More recently, though, several authors \cite{Vampa2014,Wu2015,Otobe2016} have shown that, while the intraband dynamics contributes to HHG below the band gap energy, the interband dynamics makes a main contribution to radiation above it.

To explain the mechanism of the HHG from solid-state materials, several models have been proposed.
For example,
Higuchi {\it et al.} have proposed a real-space picture using localized Wannier-Stark (WS) states and strong-field approximation, in which the differences of the quasi-energies of WS states determines the radiation energies \cite{Higuchi2014a}.
Vampa {\it et al.} have proposed a real-space three-step model analogous to its counterpart for gas-phase HHG \cite{Vampa2014,Vampa2015,Vampa2015b}.
While the pioneering works have indicated that the HHG spectra provide information about the band structure, they have considered a two-band model, with a single valence band (VB) and the first conduction band (CB).

More recently, several authors have pointed out the importance of the effects of multiple bands \cite{Hawkins2015a,Wu2015,Ndabashimiye_2016,Wu2016} (see also Ref. \cite{McDonald2015}).
Wu {\it et al.} \cite{Wu2015} have shown that the contributions from multiple bands can lead to the formation of additional plateaus, extending HHG to higher photon energies.
Ndabashimiye {\it et al.} have indeed observed the multiple-plateau harmonics in their experiment \cite{Ndabashimiye_2016} and modeled it as a dressed multi-level system \cite{Ndabashimiye_2016,Wu2016}, rather than explicitly invoking the intraband dynamics.
They have also pointed out that the dressed system can be mapped onto the band structure, which leads to a semiclassical three-step picture in momentum space.

In this paper, we show that, by drawing momentum-space electron trajectories across multiple bands, one can easily deduce many aspects of solid-state HHG such as multiple cutoff positions, time-frequency structure, and the dependence on pulse parameters.
We first simulate HHG from a one-dimensional (1D) model crystal by numerically solving the time-dependent Schr\"{o}dinger equation (TDSE).
We discretize the wave function directly on a spatial grid, as is customary for the gas phase, instead of expanding it with the Bloch or Houston basis \cite{Golde2008,Higuchi2014a,Vampa2014,Hawkins2015a,Wu2015,Vampa2015,McDonald2015,Luu2015a,Vampa2015a,Hohenleutner2015a,Wu2016}.
Thus, we automatically include the contribution from all the bands supported by the grid.
Our simulations well reproduce unique features of solid-state HHG such as the (quasi-)linear cutoff-energy scaling with the electric field strength \cite{Ghimire2011,Luu2015a} and the sudden transition from single to multiple plateaus \cite{Ndabashimiye_2016} with clear cutoffs.

Then we propose a simple model that can explain many aspects of the simulation results 
.
We trace the momentum-space electron dynamics based on interband tunneling, intraband acceleration, and recombination with the VB hole.
Once with an energy-band diagram at hand, one can apply the model without further resorting to elaborate theoretical calculations.
It should be highlighted that the electron can climb up bands by repeating interband tunneling to an upper band and intraband acceleration, based on which, our model predicts yet another difference from gas-phase HHG that the position of the highest cutoff depends on not only the wavelength and intensity of the pulse but also its duration (or number of optical cycles).

Our model can be regarded as a solid-state and momentum-space counterpart with multiband extension of the familiar trajectory analysis based on the three-step model \cite{Corkum1993,Kulander1993}, which has been highly successful for HHG from gas-phase atoms and molecules.
It provides a unified basis for understanding HHG from gaseous media and solid-state materials.
This offers a clear physical insight into the coherent electron dynamics of independent-electron nature in solids driven by intense laser field and serves as a benchmark to identify the effects of electron correlation, relaxation, dephasing, impurity, distortion, etc., in real experiments.

This paper is organized as follows.
After describing TDSE simulation methods in Sec. \ref{171727_6Nov16}, we present and discuss simulation results in Sec. \ref{171801_6Nov16}.
Then, we propose the trajectory analysis based on the solid-state three-step model in Sec. \ref{171847_6Nov16}.
Conclusions are given in Sec. \ref{171916_6Nov16}.
Atomic units are used throughout unless otherwise stated.

\section{Method}
\label{171727_6Nov16}
We consider a many-electron dynamics in a 1D model crystal along laser polarization with VBs fully occupied across the whole Brillouin zone (BZ) initially, typical of wide-band-gap semiconductors.
Within independent-electron approximation, we solve the effective TDSE for each electron in the velocity gauge:
\begin{align}
i \frac{\partial}{\partial t} \psi_{nk}(x, t) &= \hat{H}(t) \psi_{nk}(x, t)\nonumber \\
&= \left\{\frac{1}{2}\left[\frac{\nabla}{i} + A(t)\right]^2 + V(x) \right\} \psi_{nk}(x, t),
\label{eq:TDSE in general}
\end{align}
for the electron that initially lies in band $n$ with a crystal momentum $k$, where $A(t)$ is the vector potential related to the laser electric field $E(t)$ by $A(t) = - \int_{-\infty}^{t} E(t^{\prime}) dt^{\prime}$, 
and $V(x)$ the periodic single-electron effective potential of the crystal with lattice constant $a$, i.e., $V(x+a) = V(x)$.
We employ the dipole approximation, assuming that electron dynamics at macroscopically different positions are not coupled with each other \cite{Yabana2012}.
Similar 1D models have previously been used in several works \cite{Korbman2013,Higuchi2014a,Wu2015,Guan2016} and turned out to be useful.
$\psi_{nk}(x, t)$ is the time-dependent wave function whose initial state is the Bloch function $\phi_{nk}$, the eigenstate of
the field-free Hamiltonian $\hat{H}_0 = -\nabla^2/2 + V(x)$:
\begin{equation}
\hat{H}_0 \phi_{nk} = \varepsilon_{nk} \phi_{nk},
\label{eq:Bloch's theorem: Blosh functions}
\end{equation}
with $\varepsilon_{nk}$ being the energy eigenvalues.
Since the Hamiltonian retains lattice periodicity even under the action of the laser pulse, the initial crystal momentum $k$ is always a good quantum number.
Therefore, we can solve the TDSE for individual $k$ independently.

\begin{figure}[htbp]
\centering
\includegraphics[width=0.8\linewidth]{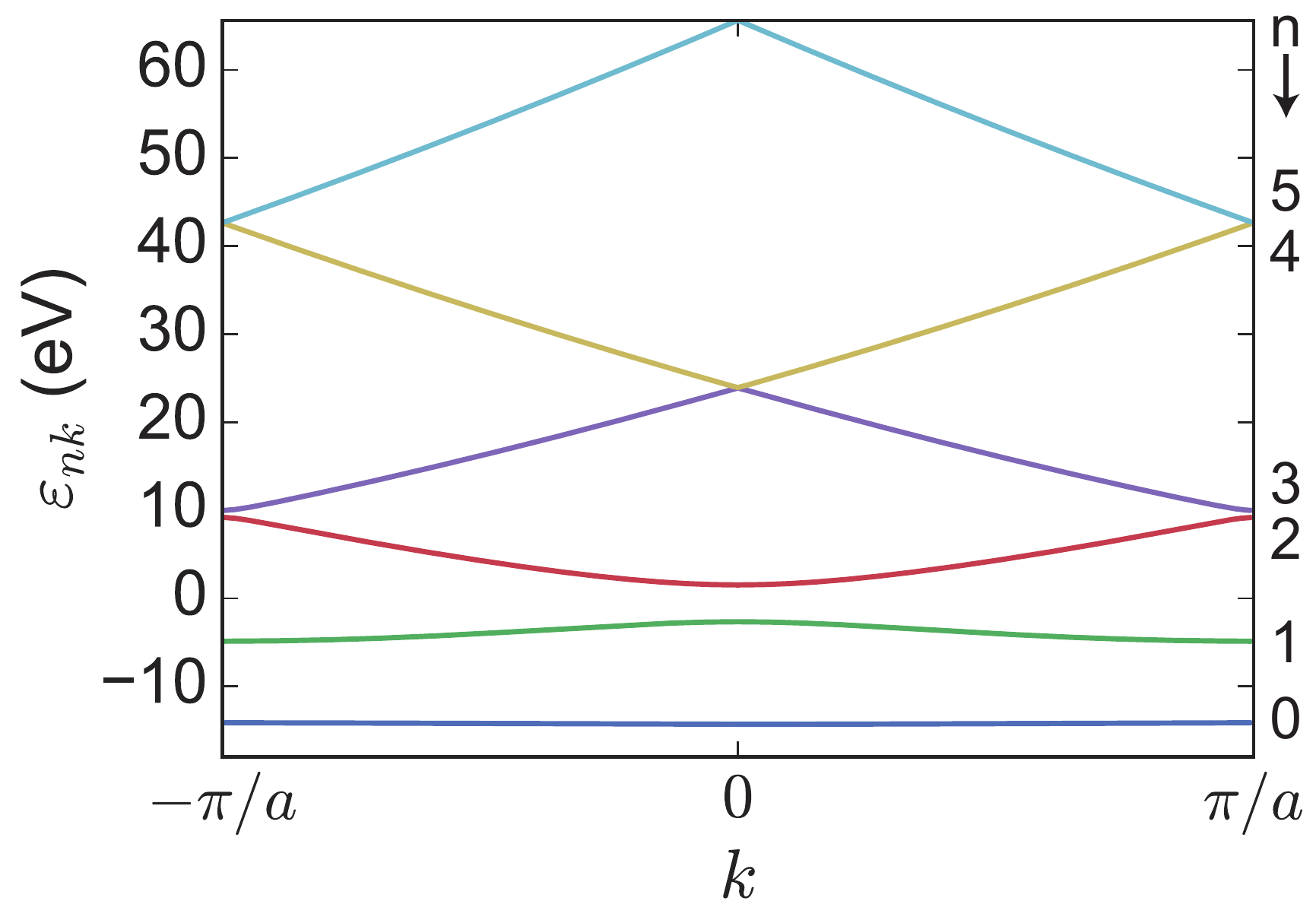}
\caption{
Two valence bands ($n=0, 1$) and first four conduction bands ($n=2, \dots, 5$) of the field-free Hamiltonian.
 The integers on the right axis are the band indices $n$.
}
\label{fig:band}
\end{figure}

Following Ref. \cite{Wu2015}, we use the Mathieu-type potential given by
\begin{equation}
V(x) = -V_0 \left[ 1 + \cos(2 \pi x / a)\right], 
\end{equation}
with $V_0 = 0.37$ and $a = 8$.
This potential expresses a band structure (Fig. \ref{fig:band}) with minimum band gap 4.2 eV at $k=0$, while the first and second CBs approach each other at the Bragg plane ($k=\pm \pi / a$).
Note that, although only six bands are shown in Fig. \ref{fig:band}, the bands taken into account in our calculation are not limited to those,
because we use the real-space basis as described below.

Instead of expanding the wave functions with basis functions \cite{Golde2008,Higuchi2014a,Vampa2014,Hawkins2015a,Wu2015,Vampa2015,McDonald2015,Luu2015a,Vampa2015a,Hohenleutner2015a,Korbman2013}, we directly solve the TDSE (\ref{eq:TDSE in general}) in real space numerically;
better convergence with the real-space basis than with the Bloch basis has previously been reported for time-dependent density-functional simulations \cite{Sato2014}.
Using Bloch's theorem, the wave function $\psi_{nk}(x, t)$ can be decomposed as,
\begin{equation}
\psi_{nk}(x,t) = e^{ikx}u_{nk}(x,t). \label{eq:Bloch's theorem} 
\end{equation}
where $u_{nk}(x,t)$ satisfies $u_{nk}(x+a,t) = u_{nk}(x, t)$.
By inserting Eq.~(\ref{eq:Bloch's theorem}) into Eq.~(\ref{eq:TDSE in general}), we obtain the equation of motion for $u_{nk}(x, t)$ as
\begin{equation}
 i \frac{\partial}{\partial t}u_{nk}(x, t) = 
  \left\{ \frac{1}{2} \left[\frac{\nabla}{i} + k + A(t) \right]^2 + V(x) \right\} u_{nk}(x, t). \label{eq:Schrodinger equation for u(x)}
\end{equation}
This is to be solved only within the unit cell $x\in [0, a]$, which enables substantial reduction of the problem size.
It should be noticed that the presence of the part $k + A(t)$ automatically accounts for the intraband dynamics \cite{Kittel1963,Krieger1986} in a natural way
and that Eq.~(\ref{eq:Schrodinger equation for u(x)}) describes interband transition among all the bands realized by the potential $V(x)$ in principle.
We assume that the two VBs ($n=0, 1$ in Fig. \ref{fig:band}) are initially filled across the whole BZ.
For a given pair of $(n, k)$, the initial Bloch functions are obtained using imaginary time propagation.
Then we numerically integrate the equation of motion (\ref{eq:Schrodinger equation for u(x)}),
using the finite difference method with the grid spacing $0.53$ a.u., time step size $2.67\times10^{-4}$ fs $= 1.10\times10^{-2}$ a.u., and the number $N$ of $k$-points 141.

\begin{figure}[tbp]
 \centering
 \includegraphics[width=0.9\linewidth]{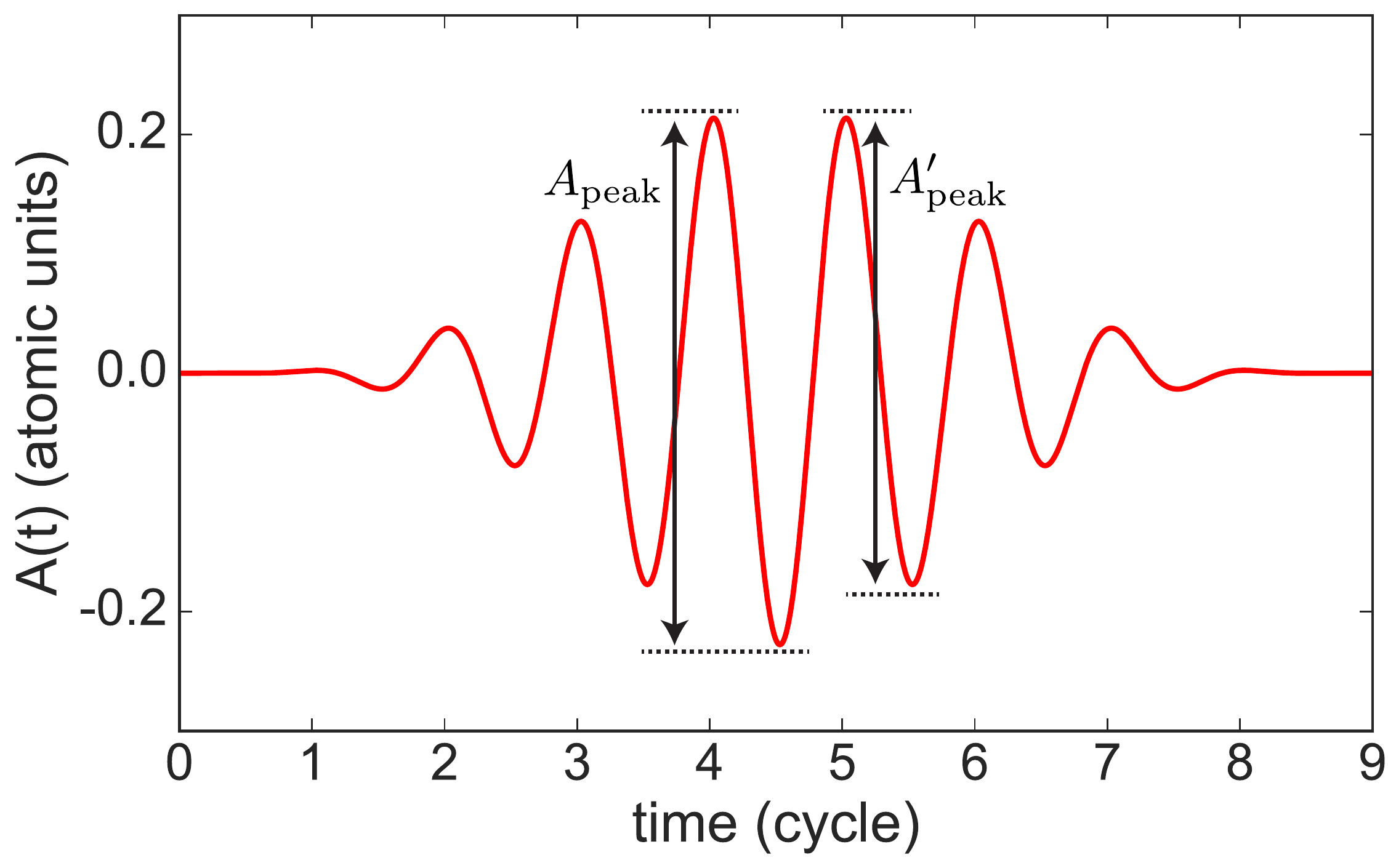}
 \caption{
 The waveform of the vector potential $A(t)$ of the laser pulse with  $E_0 = 1.65$ V/nm and $\tau = 99.66$ fs.
 The maximum and the second maximum peak-to-valley amplitude $\Apeak$ and $\Apeak^{\prime}$ are defined as depicted in the figure.
 }
 \label{fig:field and current}
\end{figure}

We calculate the contribution to the field-induced current from each $(n,k)$ as
\begin{align}
j_{nk}(t) &= \bra{\psi_{nk}(t)} \hat{p} + A(t) \ket{\psi_{nk}(t)} \nonumber \\
 &= \int_{0}^{a} u_{nk}^{\ast}(x,t) \left[ \frac{\nabla}{i} + k + A(t)\right] u_{nk}(x,t) dx.
\label{eq:each current}
\end{align}
Then $j_{nk}$ is summed over the band indices $n(=1, 2)$ and integrated over $k$ to obtain the total current
\begin{equation}
j(t) = \frac{1}{Na} \sum_{nk} j_{nk}(t).
\end{equation}
It should be remembered that $n$ and $k$ refer to the band index and crystal momentum, respectively, that the electron initially occupies.
The harmonic spectrum is calculated as the modulus square of the Fourier transform of $j(t)$.
Before applying the Fourier transform, we multiply $j(t)$ by a mask function $W(t) = \sin^4(t/\tau)$ of the same form as the field envelop in order to suppress the current after the pulse.

We consider a laser pulse of its electric field $E(t) = E_0\sin^4( t  / \tau) \sin[\omega (t - \pi\tau/2)]$ for $t \in [0, \pi \tau]$ and $E(t) = 0$ for $t \not\in[0, \pi \tau]$, where $E_0, \tau, \omega$ denote the peak electric field amplitude, a measure of pulse width, and central angular frequency, respectively (Fig. \ref{fig:field and current}).
The central angular frequency is given by $\omega = 2\pi c/\lambda$, where $c$ and $\lambda$ denote the light velocity and the central wavelength, respectively, and the central wavelength is assumed to be $\lambda = 3200$ nm.

\section{Simulation Results and Discussions}
\label{171801_6Nov16}
\begin{figure}
\includegraphics[width=0.80\linewidth]{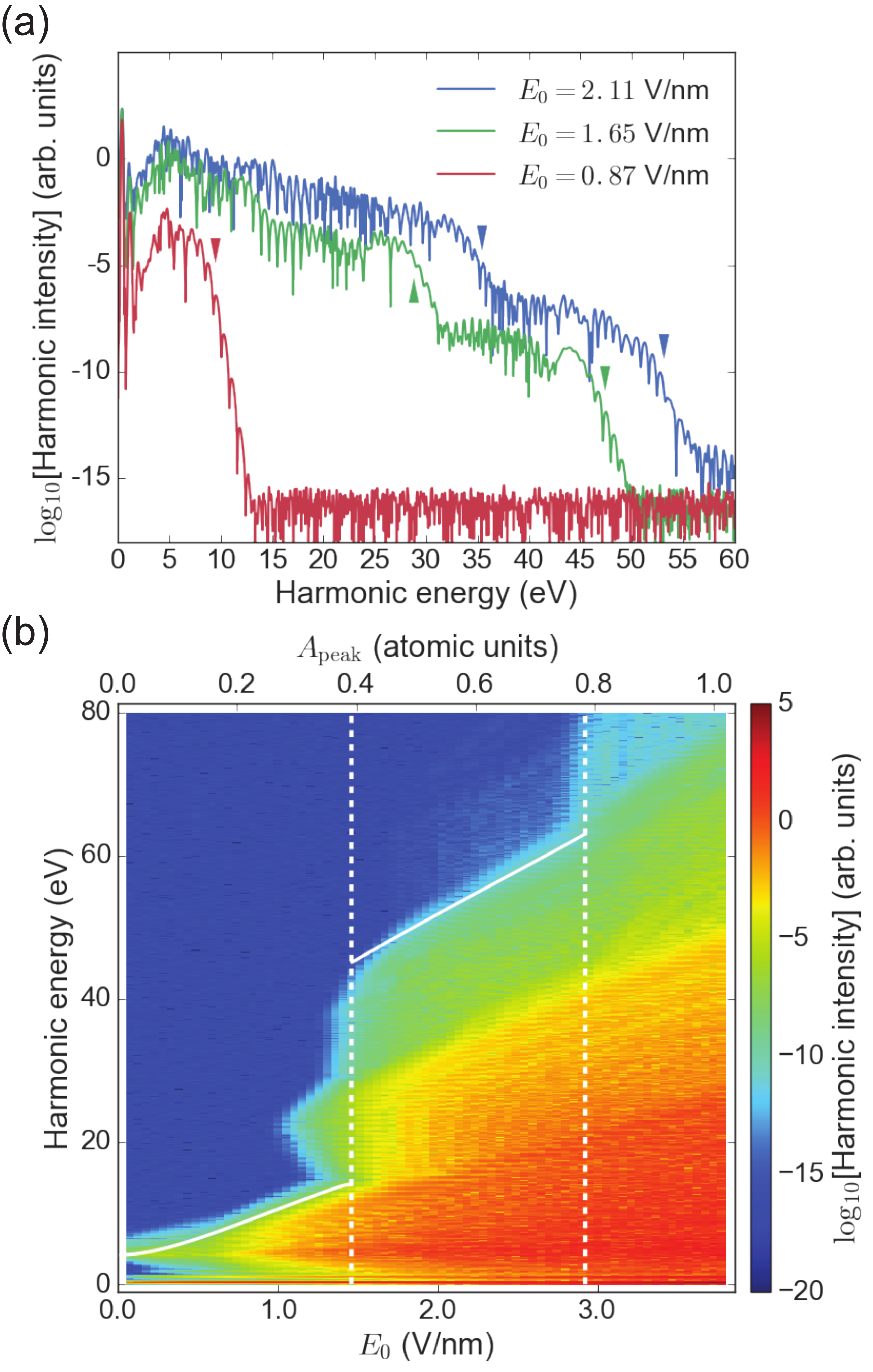}
\centering
\caption{
 (a) High harmonic spectra for $E_0 = 0.87$ V/nm (blue line), $E_0 = 1.65$ V/nm (green line), and  $E_0 = 2.11$ V/nm (red line).
 Arrowheads indicate the positions given by Eq.~(\ref{eq:cutoff 0 < Apeak < pi/a}) for $E_0 = 0.87$ V/nm (red), and Eqs.~(\ref{144641_16Nov16}) and (\ref{eq:second cutoff energy}) for $E_0 = 1.65$ (green) and $2.11$ (blue) V/nm.
 (b) False-color representation of the harmonic spectra as functions of $E_0$.
 $\Apeak$ corresponding to $E_0$ is shown on the top axis in the atomic unit.
 The two vertical white dashed lines represent $\Apeak = \pi / a$ and $\frac{2\pi}{a}$.
 The two white solid lines represent the cutoff energy positions given by Eq.~(\ref{eq:cutoff 0 < Apeak < pi/a}) for $0 < \Apeak < \frac{\pi}{a}$, and Eq.~(\ref{144641_16Nov16}) for $\frac{\pi}{a} < \Apeak < \frac{2\pi}{a}$.
}
\label{fig:harmonic spectrum}
\end{figure}

The high harmonic spectra for $\tau = 96.66$ fs, which corresponds to three cycles, are shown for several field amplitudes in Fig. \ref{fig:harmonic spectrum} (a).
While the spectrum for $E_0 = 0.87$ V/nm has a single plateau and cutoff similarly to atomic HHG,
those for $E_0 = 1.65$ and $2.11$ V/nm have two additional plateaus of lower intensity, for example, for $E_0=1.65$ V/nm, the second plateau lies at $\approx$ 15-30 eV and the third $\approx$ 30-50 eV.
In Fig. \ref{fig:harmonic spectrum}(b), we show the harmonic spectra as functions of $E_0$ (bottom axis).
The transition from the single- to multiple-plateau structure takes place not gradually but suddenly at $E_0 \approx 1.4$ V/nm.
Thus, our simulations reproduce the unique features of solid-state HHG previously reported both theoretically and experimentally \cite{McDonald2015,Wu2015,Ndabashimiye_2016}.

Let us now take a closer look at Fig. \ref{fig:harmonic spectrum}(b).
While the cutoff energy increases smoothly with $E_0$ up to 1.4 V/nm, second and third plateaus suddenly appear, and the cutoff jumps up from 15 eV to 45 eV at $E_0 \approx 1.4$ V/nm.
Moreover, another cutoff jump is seen at $E_0 \approx$ 2.8 V/nm, from 60 eV.
If we let $\Apeak$ denote the maximum peak-to-valley amplitude of $A(t)$ (see Fig.~\ref{fig:field and current}) and show it on the top axis of Fig.~\ref{fig:harmonic spectrum}(b), we notice that, interestingly, the
jump-up positions well coincide with the condition that $\Apeak = \frac{\pi}{a} = 0.393$ a.u. and $\frac{2\pi}{a} = 0.786$ a.u. [vertical white dashed lines in Fig. \ref{fig:harmonic spectrum}(b)].
Note that $\frac{\pi}{a}$ is the distance from the $\Gamma$ point to the first-BZ edge (Fig.~\ref{fig:band}).
Although $\Apeak$ may be approximated by $2A_0$, with $A_0$ being the amplitude of the vector potential, in many practical situations, we use $\Apeak$ in the present study, since it directly characterizes the largest crystal momentum gain in the intraband dynamics, as we will see in the next section.

Whereas the first cutoff at $0 < \Apeak < \pi / a$ and the third at $\pi / a < \Apeak < 2\pi / a$ appear to increase quasi-linearly with the field strength \cite{Ghimire2011,Luu2015a},
it seems that they are closely related with the particle-hole energy, defined as,
\begin{equation}
\varepsilon_{nm}(k):= \varepsilon_{nk}-\varepsilon_{mk},
\end{equation}
between bands $m$ and $n$ at a crystal momentum $k$ \footnote{
Given that the band structure is periodic with reciprocal lattice vector $\frac{2\pi}{a}$, we allow $k(t)$ beyond the first BZ.
Alternatively, one can confine it to the first BZ by rewriting, e.g., Eq.~(\ref{144623_16Nov16}) as $E_{31} = \Delta\varepsilon_{31}(\Apeak - \frac{2\pi}{a})$.
}.
At $\Apeak < \frac{\pi}{a}$, the cutoff energy agrees well with
\begin{equation}
 \Delta \varepsilon_{21}(\Apeak), \label{eq:cutoff 0 < Apeak < pi/a}
\end{equation}
[white solid line at $E_0 < 1.4$ V/nm in Fig. \ref{fig:harmonic spectrum}(b)],
consistent with the formula proposed by Vampa {\it et al.} \cite{Vampa2015} for a two-band system.
The multiple cutoff positions at $\Apeak > \frac{\pi}{a}$, on the other hand, cannot be explained by Eq.~(\ref{eq:cutoff 0 < Apeak < pi/a}).

\section{Trajectory analysis}
\label{171847_6Nov16}

In this section, we propose a simple model to explain the above findings as well as cutoff positions for $\Apeak > \frac{\pi}{a}$ and the time-frequency structure of HHG.
Its essential ingredients are summarized as follows:
\begin{enumerate}
 \renewcommand{\labelenumi}{(\roman{enumi})}
 \item Each electron is tunnel ionized to an upper band predominantly at the minimum band gap, e.g., from band 1 to 2 at $k=0$ and from 2 to 3 at the BZ edge.
 \item the laser-driven intraband dynamics is expressed by displacement in the momentum space as $k(t) = k_0 + A(t)$ where $k_0$ denotes the initial crystal momentum
       \footnote{In situations relevant with HHG, $A(t)$ is comparable with or even larger than the BZ width.}
       (this is known as the acceleration theorem \cite{Kittel1963,Krieger1986}).
 \item Each electron emits a photon when it undergoes an interband transition to the initial band.
       The photon energy is given by the particle-hole energy $\Delta\varepsilon_{n(t)n_0}[k(t)]$ between the band $n(t)$ where the electron is located at $t$ and the initial band $n_0$.
\end{enumerate}

\begin{figure}
 \centering
 \includegraphics[width=1.00\linewidth]{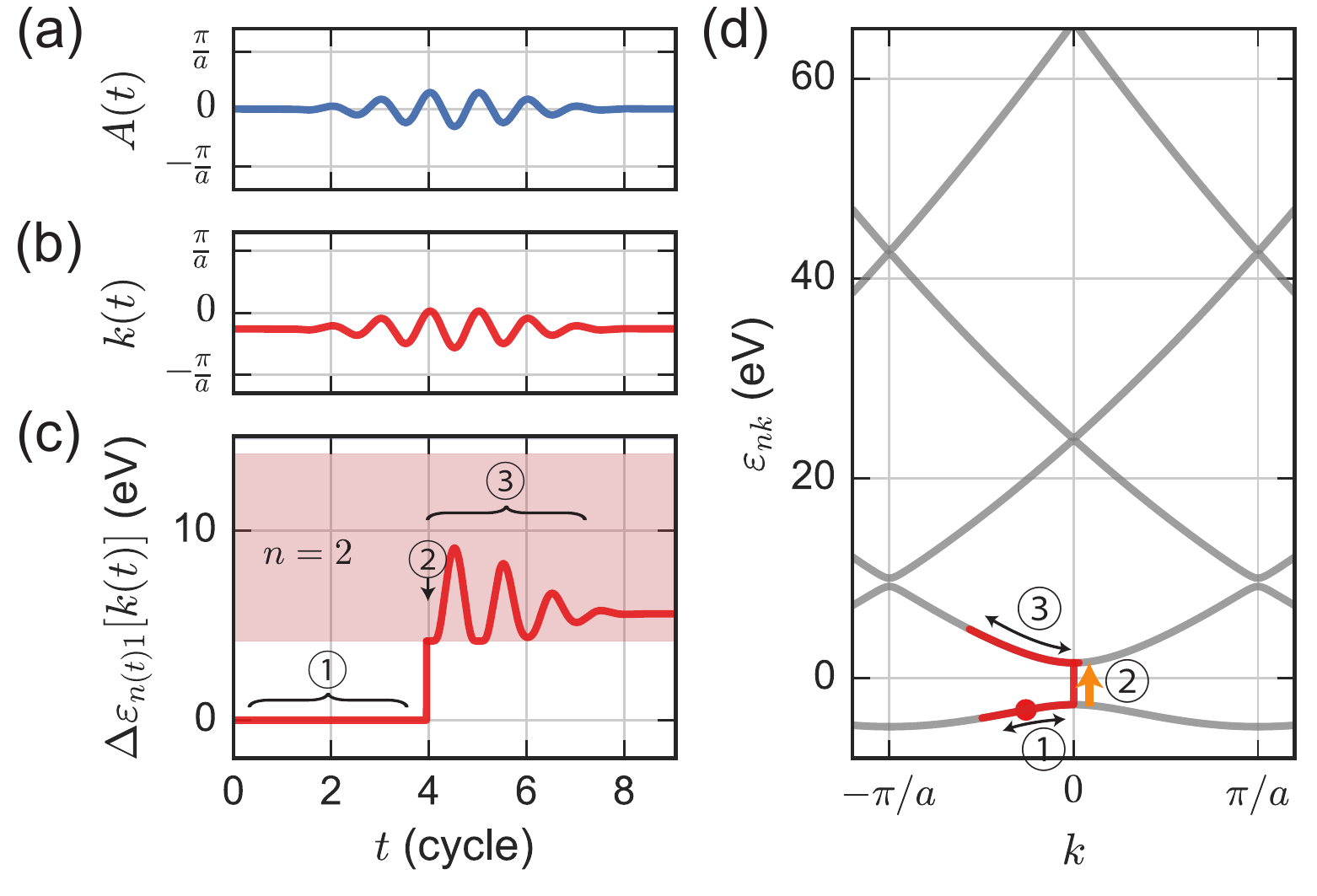}
 \caption{
 Momentum-space trajectory of an electron excited from a VB ($n=1$) to the first CB ($n=2$) at $t=4$ cycles, drawn based on the solid-state three-step model when $E_0 = 0.87$ V/nm or $\Apeak = 0.23 < \frac{\pi}{a}$, for which $k_0 = -0.26 \times \frac{\pi}{a}$.
 (a) waveform of $A(t)$
 (b) instantaneous crystal momentum $k(t)$
 (c) temporal evolution of the particle-hole energy, i.e., emitted photon energy
 (d) pictorial representation of the momentum-space electron trajectory in the band diagram.
 }
 \label{fig:pictorial_weak}
\end{figure}

Interestingly, (i)-(iii) are reminiscent of tunneling ionization, acceleration, and recombination, respectively, in the three-step model \cite{Corkum1993,Kulander1993} of gas-phase HHG.
Whereas such an analogy has been suggested also in Refs. \cite{Vampa2014,Vampa2015}, several remarks are in order:
\begin{itemize}
 \item Our solid-state three-step model follows electron dynamics (and trajectories) in the momentum space whereas the gas-phase three-step model considers it in the coordinate space;
       the momentum-space analysis is more natural and convenient for Bloch electrons in a periodic potential.
 \item All the electrons in the VB undergo the intraband acceleration (ii) together \cite{Kittel2004,Ashcroft1976} even before the first tunneling.
       Thus, VB electrons starting from not only $k_0=0$ (as assumed in Refs. \cite{Wu2015,Ndabashimiye_2016}) but also any arbitrary values of $k_0$ are considered \footnote{This does not violate the Pauli exclusion principle, since all the electrons in the VB move uniformly together \cite{Kittel2004,Ashcroft1976}, and thus, no $(n, k)$ point is occupied simultaneously by more than one electron at any time.}.
 \item The electron can climb up to higher and higher bands  by repeating (i) and (ii).
 \item Unlike in the gas phase, (ii) also contributes to harmonic generation \cite{Ghimire2012,Hawkins2013,Vampa2014,Luu2015a,Wu2015}.
 \item (iii) can take place at any time in a trajectory, in contrast to the atomic case where the electron can recombine with the parent ion only when it returns to the nuclear position.
\end{itemize}

\begin{figure}
 \centering
 \includegraphics[width=1.00\linewidth]{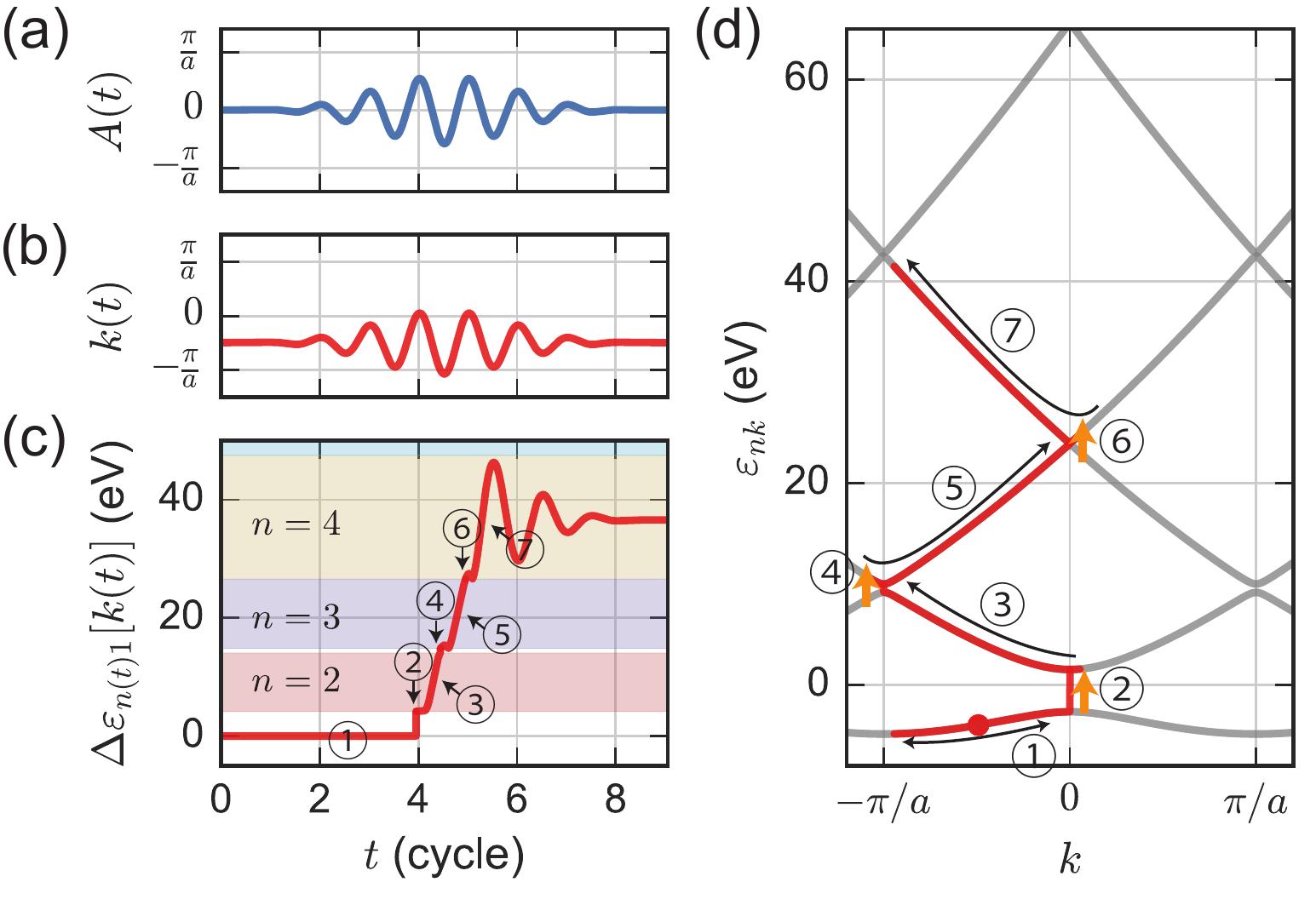}
 \caption{
  Same as Fig.~\ref{fig:pictorial_weak}, but when $E_0 = 1.65$ V/nm or $\Apeak = 0.44 > \frac{\pi}{a}$, for which $k_0 = -0.49 \times \frac{\pi}{a}$.
 }
 \label{fig:pictorial-strong}
\end{figure}

In the case of gas-phase three-step model, one can trace a classical electron trajectory in the coordinate space for each ionization time, which explains the cutoff law and the time-frequency structure.
In the solid case, analogously, using the above-mentioned recipes, we can trace an electron trajectory in the band diagram for each time $t_0$ of interband tunneling at the $\Gamma$ point from a VB to a CB.
We present an example when $\Apeak = 0.23 < \frac{\pi}{a}$ ($\Apeak = 0.44 > \frac{\pi}{a}$) in Fig.~\ref{fig:pictorial_weak} (Fig.~\ref{fig:pictorial-strong}).
Note that, once given a waveform of $A(t)$ [Fig.~\ref{fig:pictorial_weak}(a) and Fig.~\ref{fig:pictorial-strong}(a)], the crystal momentum displacement can be fully described as $k(t) = k_0 + A(t)$ with $k_0 = -A(t_0)$ as shown in Fig.~\ref{fig:pictorial_weak}(b) and Fig.~\ref{fig:pictorial-strong}(b).

First, we discuss the electron dynamics when $\Apeak < \frac{\pi}{a}$ [Fig. \ref{fig:pictorial_weak}(d)].
Electrons initially in the valence band are accelerated (\textcircled{\scriptsize 1}),
and excited to the CB at $k=0$ at $t=t_0$ (\textcircled{\scriptsize 2}).
The subsequent momentum change is given by
\begin{equation}
 k(t) = k_0 + A(t) = A(t) - A(t_0), \label{eq:momentum change in CB}
\end{equation}
and hence, $|k(t)| < \Apeak$.
Thus, the maximum displacement in the first CB is $\Apeak$.
Now that $\Apeak <  \pi / a$, no electrons can reach the BZ edge, but they oscillate in the first CB without further excitation (\textcircled{\scriptsize 3}).
Hence, the emitted photon energy is given as a function of recombination time $t$ by $\Delta\varepsilon_{21}[k(t)]$ [Fig.~\ref{fig:pictorial_weak}(c)], and, the highest energy of the photon 
is given by 
\begin{equation}
 \Delta \varepsilon_{21}(\Apeak),
\end{equation}
which agrees  with the cutoff position in Fig. \ref{fig:harmonic spectrum}(b) and Eq.~(\ref{eq:cutoff 0 < Apeak < pi/a}).

\begin{figure*}
 \includegraphics[width=0.85\linewidth]{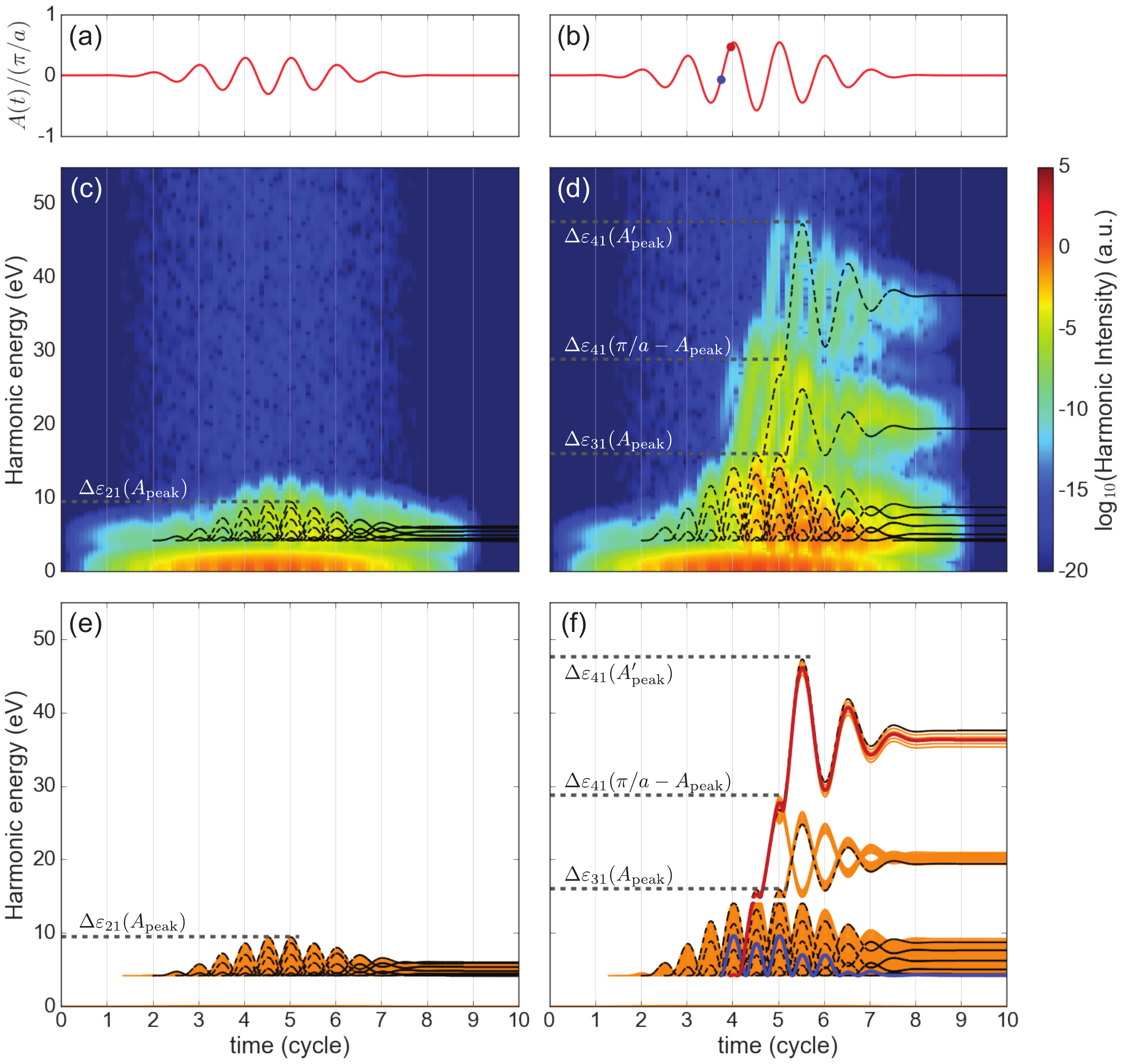}
 \centering
 \caption{
 The temporal evolution of high harmonic generation.
 (a), (c), and (e) are for $E_0 = 0.87$ V/nm or $\Apeak = 0.23 < \frac{\pi}{a}$, while (b), (d), and (f) are for $E_0 = 1.65$ V/nm or $\Apeak = 0.44 > \frac{\pi}{a}$. 
 (a) and (b):vector potential normalized to $\pi / a$.
 (c) and (d):time-frequency analysis of HHG. 
 Gabor transform of the TDSE simulation results with a temporal window having a FWHM of $1.78$ fs, i.e., $2.3$ eV in energy.
 (e) and (f): orange solid lines denote electron energy trajectories that first tunnel from the VB to the first CB at $k=0$ at different momenta (not necessarily at the first approach) and then climb to upper bands as soon as the electron reaches the minimum band gaps.
 The blue line in (f) represents the trajectory of an electron which is first tunnel-ionized at $A(t) \sim 0$ [blue circle in (b)], whereas the red line in the vicinity of the peak of $A(t)$ [red circle in (b)].
 In (c) and (d) we show by black dashed lines the trajectories corresponding to the black dashed lines in (e) and (f), respectively, to facilitate comparison.
 The horizontal gray dashed lines in (c), (d), (e), and (f) show the predicted cutoff energy positions, i.e., $\Delta \varepsilon_{21}(\Apeak)$ for (c) and (e) ($\Apeak = 0.23$), and $\Delta \varepsilon_{31}(\Apeak)$, $\Delta \varepsilon_{41}(\frac{\pi}{a} - \Apeak)$, and $\Delta \varepsilon_{41}(\Apeak^{\prime})$ from the bottom for (d) and (f) ($\Apeak = 0.44$), respectively.
}
 \label{fig:temporal harmonic}
\end{figure*}

Next, let us turn to the case $\Apeak > \pi / a$ [Fig. \ref{fig:pictorial-strong}(d)].
After excitation to the first CB (\textcircled{\scriptsize 1} - \textcircled{\scriptsize 2}), part of electrons can now be accelerated to reach the BZ edge (\textcircled{\scriptsize 3}),
and open a channel to climb up to the upper CB (\textcircled{\scriptsize 4}) within a half cycle.
The promoted electrons then undergo intraband displacement to the reversed direction in the second CB ($n=3$) in the next half cycle, enabling photon emission of higher energy (\textcircled{\scriptsize 5}), which neatly explains why multiple plateaus appear at $\Apeak \approx \frac{\pi}{a}$ [Fig.~\ref{fig:harmonic spectrum}(b)].
Whereas we have assumed interband transitions precisely at the minimum band gaps, they can also take place in their vicinities in reality.
This explains the appearance of some high-energy components even before $\Apeak$ reaches $\frac{\pi}{a}$ in Fig.~\ref{fig:harmonic spectrum}(b), from $E_0 \sim 1.1$ V/nm.

Every time the electrons reach the minimum energy gap to next CB each half cycle, they can undergo successive interband excitation (\textcircled{\scriptsize 5} - \textcircled{\scriptsize 7}) (or pass through it).
If the second maximum peak-to-valley amplitude is denoted by $\Apeak^{\prime}$ (Fig. \ref{fig:field and current}), they can climb up to the third CB ($n=4$) if $\Apeak^\prime < \pi / a$ and the fourth CB ($n=5$) if $\Apeak^\prime > \pi / a$ at $t\approx 5.5 T$ with $T$ being the optical cycle.
From this scenario, we can estimate the maximum energy gain as
\begin{align}
 E_{c} = \left\{
\begin{array}{lc}	
\Delta \varepsilon_{41}(\Apeak^{\prime}) & (\Apeak^{\prime} < \frac{\pi}{a}) \\
\Delta \varepsilon_{51}(\Apeak^{\prime}) & (\frac{\pi}{a} < \Apeak^{\prime}),
\end{array}
 \right.
 \label{144641_16Nov16}
\end{align}
which reproduces the highest harmonic energy in Fig.~\ref{fig:harmonic spectrum}(b).
It should be noted that the highest cutoff energy can exceed that in the gas phase for the same laser parameters and ionization potential (band gap energy in the solid case), as has been recently observed \cite{Ndabashimiye_2016}.

The temporal profile of the photon energy emitted from the trajectory in Fig.~\ref{fig:pictorial_weak}(d) [Fig.~\ref{fig:pictorial-strong}(d)] is given by Fig.~\ref{fig:pictorial_weak}(c) [Fig.~\ref{fig:pictorial-strong}(c)], since the electron can recombine with the VB hole at any time, as prescribed.
By accumulating similar curves for all possible values of $t_0$ and climb-up-or-pass-through branchings, one can deduce the time-frequency structure of HHG, as displayed in Fig.~\ref{fig:temporal harmonic}(e) and (f).
They indeed
capture the main features of the HHG temporal structure, extracted from The TDSE simulation results through Gabor transformation, above the minimum band gap (4.2 eV) [Fig. \ref{fig:temporal harmonic}(c) and (d)].
The below-band-gap harmonics are emitted through the intraband dynamics.

The temporal structure under $\Apeak < \frac{\pi}{a}$ shown in Fig. \ref{fig:temporal harmonic}(c) is consistent with that previously discussed by Vampa {\it et al.} \cite{Vampa2015}.
It is also noteworthy that this electron dynamics is conceptually similar to that in harmonic generation from graphene \cite{Ishikawa2010,Ishikawa2013,Bowlan2014}.
For $\Apeak > \frac{\pi}{a}$, in contrast, Fig.~\ref{fig:temporal harmonic}(d) and (f) contain step-like features, stemming from the band-climbing process \footnote{This somewhat reminds us of {\it Donkey Kong}, an arcade game released by Nintendo (\url{https://en.wikipedia.org/wiki/Donkey_Kong_(video_game)})} unique to solid-state materials;
they manifest themselves as multiple plateaus and cutoffs seen in Fig.~\ref{fig:harmonic spectrum}.
Some step heights indicated with the horizontal dashed lines in Fig.~\ref{fig:temporal harmonic}(f) are characterized by
\begin{align}
 E_{31} = &\Delta \varepsilon_{31}(\Apeak),\label{144623_16Nov16} \\
 E_{41} = &\Delta \varepsilon_{41}(\frac{\pi}{a} - \Apeak),\label{eq:second cutoff energy}
\end{align}
which well agree with the TDSE simulation results [horizontal dashed lines in Fig.~\ref{fig:temporal harmonic}(d) and arrowheads in Fig.~\ref{fig:harmonic spectrum}(a)].
The remaining difference between Fig.~\ref{fig:temporal harmonic}(d) and (f) can be accounted for again by interband transition not just precisely at the minimum band gap but also in its vicinity.
Note that $-\frac{\pi}{a} < \frac{\pi}{a} - \Apeak < 0$, and therefore, that $E_{41}$ increases with increasing $\Apeak$.

Electrons starting from $k_0 \sim 0$ are excited when $A(t) \approx 0$, or, at an extremum of $E(t)$ [the blue circle in Fig.~\ref{fig:temporal harmonic}(b)], which favors tunneling transition.
On this basis, one might argue that they would make a main contribution to HHG \cite{Vampa2015}.
It should, however, be noticed that they cannot reach the BZ edge and are confined in the first CB [blue line in Fig.~\ref{fig:temporal harmonic}(f)] unless $\Apeak \le \frac{2\pi}{a}$.
As a consequence, their contributions are limited to the range below $E_{31}$.
In contrast, the harmonic components above $E_{31}$ including the highest cutoff are dominated by the electrons [red line in Fig.~\ref{fig:temporal harmonic}(f)] that are initially far from the $\Gamma$ point and
first excited in the vicinity of a peak of $A(t)$ [the red circle in Fig. \ref{fig:temporal harmonic}(b)] or $E(t) \approx 0$, thus with smaller probability.
This may be one of the reasons why higher plateaus are weaker in intensity.

\begin{figure}[tbp]
 \centering
 \includegraphics[width=0.98\linewidth]{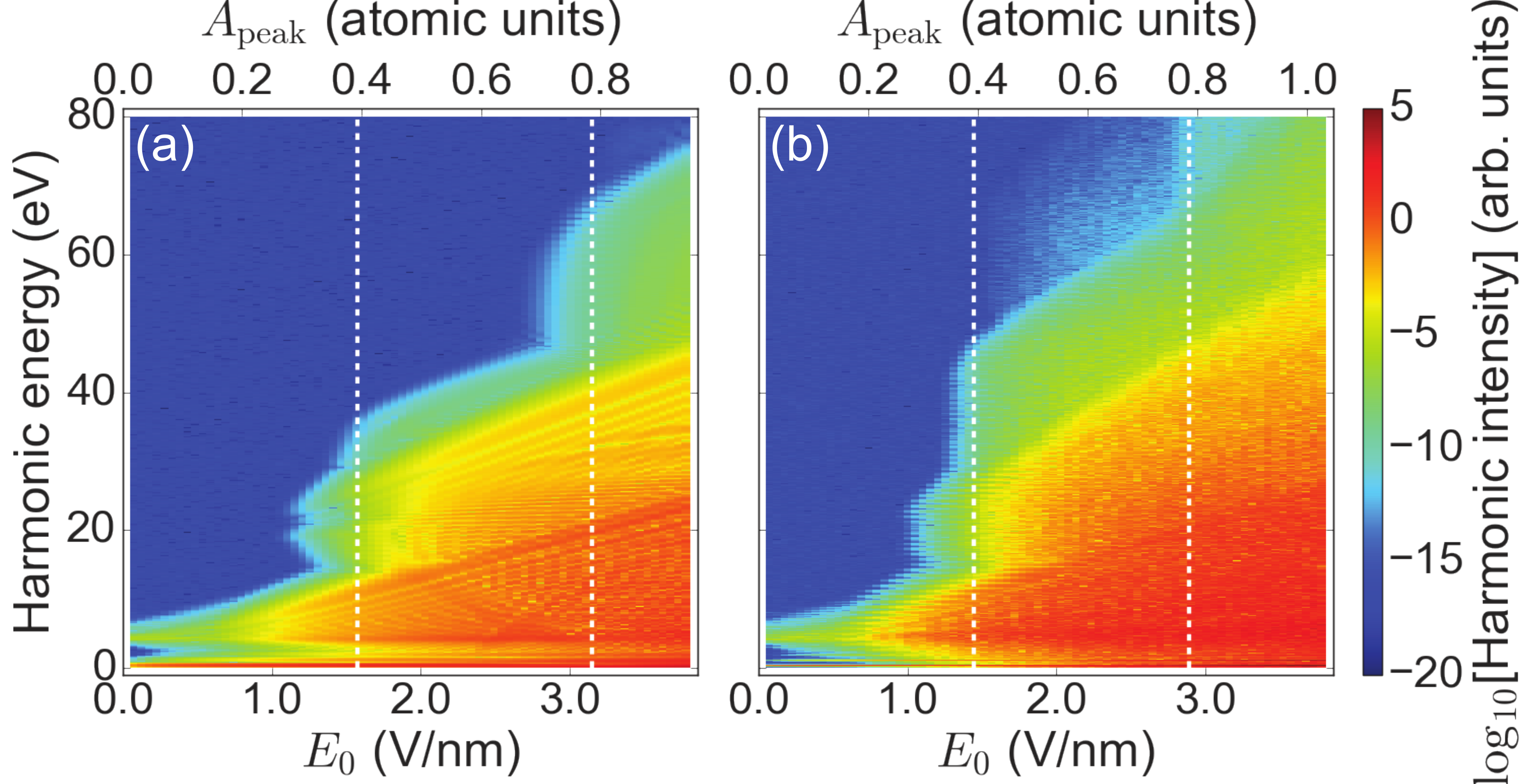}
 \caption{
 False-color representation of HHG spectra as functions of $E_0$ for (a) single-cycle driving field ($\tau = 42.70$ fs) and (b) multi-cycle field ($\tau = 192.14$ fs).
The corresponding $\Apeak$ is shown on the top axis in the atomic unit.
The vertical white dashed lines represent $\Apeak = \frac{\pi}{a}$ and $2\pi / a$.
 }
 \label{fig:various cycle harmonics}
\end{figure}

An intriguing prediction of the present model is that the number of plateaus and the highest cutoff energy depend not only on wavelength and electric field amplitude (or vector potential amplitude) but also on pulse width or number of cycles, in marked contrast to gas-phase HHG.
This is confirmed by Fig. \ref{fig:various cycle harmonics}, which compares harmonic spectra for single-cycle ($\tau = 42.70$ fs) and multi-cycle ($\tau = 192.14$ fs) driving fields.
One can clearly see that the third plateau is missing at $0.39 < \Apeak < 0.79$ for the shorter pulse.

\section{Conclusions}
\label{171916_6Nov16}
We have proposed a simple model to describe HHG spectra from periodic crystalline solids, based on intraband displacement driven by the vector potential, tunneling between multiple bands, and interband recombination to the valence band.
Our model can be viewed as a solid-state and momentum-space counterpart of the familiar three-step model for the gas phase \cite{Corkum1993,Kulander1993}.
If the intraband dynamics allows the electron to reach the BZ edge, repeated tunneling and intraband displacement lead to multiple plateaus, which is one of the recently observed unique features of solid-state HHG \cite{Wu2015,McDonald2015,Ndabashimiye_2016}.
Our model can successfully reproduce the laser intensity at which the multiple-plateau structure appears, cutoff energy positions, and temporal structure of HHG calculated through numerical solution of the single-electron TDSE.
Moreover,
it predicts that the cutoff energy depends on not only laser intensity and wavelength but also pulse width.
Expectedly, one can further refine the present model by incorporating an interband tunneling rate dependent on $(n, k)$.

It may be useful to briefly mention the similarity and difference between our model and the recently proposed model of a strongly driven (or dressed) multi-level system \cite{Ndabashimiye_2016,Wu2016}.
Assuming that the dressed state ultimately reproduces the band structure, their model appears to describe the physics similar to that in our model, in principle.
It should be, however, emphasized that whereas their model considers the contribution only from the VB electron initially located at $k=0$, we properly take the contribution from all the VB electrons into account.
Moreover, by treating the intraband dynamics explicitly as the crystal momentum displacement induced by the vector potential of the laser pulse, our model can directly connect the emergence of multiple plateaus and cutoff energies with the band structure in a clear-cut manner.

Thus, our model will offer a new way to investigate and control the electronic state in solid materials with intense laser fields, such as the reconstruction of band structure from high-harmonic spectra and control of excited electron population via waveform.

\begin{acknowledgments}
We thank Kuniaki Konishi for helpful discussions.
K. L. I. thanks Mette Gaarde for meaningful discussions.
This research is supported in part by Grants-in-Aid for Scientific Research (No. 25286064, No. 26390076, No. 26600111, and No. 16H03881) from the Ministry of Education, Culture, Sports, Science and Technology (MEXT) of Japan, and also by the Advanced Photon Science Alliance (APSA) project commissioned by MEXT. 
This research is also partially supported by the Photon Frontier Network Program of MEXT, 
by the Center of Innovation Program from the Japan Science and Technology Agency, JST, 
by Core Research for Evolutional Science and Technology, Japan Science and Technology Agency (CREST, JST),
and by the MEXT as ``Exploratory Challenge on Post-K computer''.
T. I. was supported by a JSPS Research Fellowship.
\end{acknowledgments}

\bibliographystyle{apsrev4-1}
\bibliography{reference}

\end{document}